\documentclass[aps,prb,twocolumn,showpacs,groupedaddress]{revtex4}
\usepackage{graphicx}

\begin{document}


\title{Bistable Spin Currents from Quantum Dots Embedded in a Microcavity}

\author{Ivana Djuric and Chris P. Search}
\affiliation{Department of Physics and Engineering Physics,
Stevens Institute of Technology, Hoboken, NJ 07030}

\date{\today}

\begin{abstract}
We examine the spin current generated by quantum dots embedded in
an optical microcavity. The dots are connected to leads, which
allow electrons to tunnel into and out of the dot. The spin
current is generated by spin flip transitions induced by a
quantized electromagnetic field inside the cavity with one of the
Zeeman states lying below the Fermi level of the leads and the
other above. In the limit of strong Coulomb blockade, this model
is analogous to the Jaynes-Cummings model in quantum optics. We
find that the cavity field amplitude and the spin current exhibit
bistability as a function of the laser amplitude, which is driving
the cavity mode. Even in the limit of a single dot, the spin
current and the Q-distribution of the cavity field have a bimodal
structure.
\end{abstract}

\pacs{42.50.Pq,73.63.Kv,78.67.Hc} \maketitle

\section{Introduction}
Optical bistability (OB) occurs when a nonlinear dielectric is
placed inside of optical resonator, which provides a feedback
mechanism for the light. The output intensity from the resonator
can, as a result, have two or more stable values for a given input
intensity that is driving the cavity and can be made to switch
between these two outputs by varying the input intensity beyond
the bistable region\cite{Meystre}. The interest in OB systems
started in the 1960's with $Sz\ddot{o}ke's$ et al theory of
absorptive optical bistability where the bistability is a result
of absorption by the dielectric medium\cite{Szoke}.  OB was first
observed experimentally and explained by Gibbs {\em et al.}
\cite{Gibbs} for a cavity containing a medium with a nonlinear
index of refraction and no absorption or gain (dispersive
bistability). Interest in OB has been stimulated by its practical
applications including optical switches, logic gates, and
optically bistable memory devices\cite{Miller, Abraham2, Gibbs2,
Mandel,Waren}. In addition to device applications, OB is also
interesting because it exhibits novel physical behavior such as
phase transitions between stationary but non-equilibrium
states\cite{Abraham,Bonifacio}.

In a completely independent development, spintronics has emerged
as new field in which the spin degrees of freedom of charge
carriers in solid state devices are exploited for the purpose of
information processing. Manipulation of the spin degrees of
freedom rather than the charge has the advantage of longer
coherence and relaxation times \cite{zutic}. In order to
manipulate the spin degree of freedom for the purpose of
information processing, there is a demand for efficient and
readily fabricated spin devices such as spin batteries, spin
filters, spin transistors, etc.. Much of this work has focused on
ways to generate pure spin currents in semiconductor
nanostructures using, for example, the extrinsic Spin-Orbit (SO)
interaction \cite{extrinsic_SO}, Rashba SO interactions
\cite{rashba}, optical absorption \cite{stevens}, Raman scattering
\cite{najmaie}, shape deformations of open quantum dots
\cite{mucciolo,watson}, as well as various types of quantum pumps
\cite{sharma,benjamin,blaauboer,sela}.  Pure spin currents,
$I_s=s(I_{\uparrow}-I_{\downarrow}),$ are the result of an equal
number of spin up ($\uparrow$) and spin down ($\downarrow$) charge
carriers moving in the opposite direction so that the charge
current is zero, $I_c=q(I_{\uparrow}+I_{\downarrow})$. Here,
$I_{\sigma}$ are the spin polarized particle currents, $s=\hbar/2$
the spin of the particle, and $q=e$ the charge.

Another model for a spin battery that has recently been proposed
is electron spin resonance (ESR) in a quantum dot connected to
leads, which generates a pure spin current by spin flip
transitions when there is a large Zeeman splitting
\cite{wang-zhang,dong}. A classical transverse magnetic field was
used to induce the spin flips and hence create the spin current.
We recently extended this model by considering spin flips induced
by a {\em quantized} mode of an optical
microcavity\cite{Djuric-Search-1}. In this case, a two-photon
Raman transition via an intermediate charged exciton (trion) state
was used to induce spin flips. The spin current was found to be
significantly larger in our case than for a classical undepleted
field as a result of the cavity decay. Also, the shot noise
exhibited a rich structure that was consequence of the discrete
photon numbers in the cavity.

In our previous work, a spin flip from the lower to upper Zeeman
states involved the absorption of a photon from a classical pump
laser and creation of photon in the cavity mode. As a result, the
cavity field was built up out of the vacuum without any need for
driving the cavity \cite{Djuric-Search-1}. Here, we consider the
reverse process in which a photon from the cavity mode must be
absorbed in order to flip the electron spin. This requires that
the cavity mode now be driven by an external source.

We show here that the driven cavity system exhibits absorptive OB
for the amplitude of the cavity field. Because the spin current is
a function of the cavity field amplitude, the spin current also
exhibits bistability as function of the amplitude of the driving
field. We study the limit of both a single quantum dot coupled to
the cavity as well as $N\gg 1$ quantum dots coupled to the cavity.
In the later case, we can use a semiclassical treatment for the
dots and cavity field \cite{armen}. This behavior indicates that
the system could be used as an optically controlled spin current
switch. While optical bistability has been studied in quantum
wells embedded in semiconductor microcavities \cite{gurioli}, this
is the first study of OB in the presence of electrical transport
through quantum dots.

In Section II, we outline our model. In Section III, we analyze
the steady state solution of the master equation for a single
quantum dot coupled to the cavity. In Section IV, we consider the
semiclassical solution for $N\gg 1$ dots coupled to the cavity
mode. We examine the effect that inhomogeneous broadening of the
dots' Zeeman splitting and variations of dots' vacuum Rabi
frequency have on the bistability.

\section{Model}
We consider self-assembled quantum dots embedded in a high-Q
microcavity, as depicted in Fig. 1. Strong coupling between
individual self-assembled and interface fluctuation quantum dots
with a single mode of an optical microcavity has recently been
achieved \cite{reithmaier,peter}. Here we are interested in
simultaneous coupling of dots to a cavity mode and electrical
transport through the dots due to tunnelling from a doped
reservoir. Although self-assembled quantum dots are usually used
for optical studies, there have been several experimental studies
of electrical transport through individual self-assembled InAs
quantum dots \cite{schmidt,ota,barthold} as well as through a thin
sheet containing a large number of InAs dots \cite{kieblich}.
Along a similar line, the ability to control the tunnelling of
electrons or holes between self-assembled dots and a doped GaAs
reservoir by a gate voltage combined with simultaneous
spectroscopic studies of these charged quantum dots has been
demonstrated \cite{petroff,atature}.

\begin{figure}[htb]
\includegraphics[height=3.5 in]{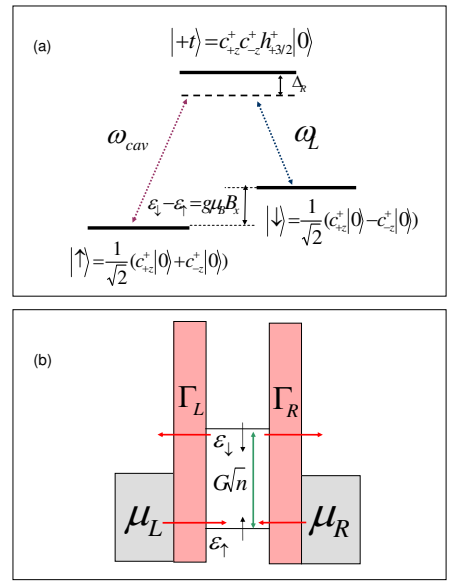}
\caption{(a) Raman transition between the dot Zeeman states,
$|\uparrow,\downarrow \rangle$, via an intermediate trion state,
$|+t\rangle$, induced by a laser with frequency $\omega_L$ and a
cavity mode with frequency $\omega_{cav}$. The spin eigenstates
along the direction of the magnetic field are superpositions of
spin eigenstates in the growth direction, $\hat{c}^{\dagger}_{\pm
z}|0\rangle$. (b) Schematic of a single quantum dot indicating
Zeeman energy levels in the dot and allowed tunnelling between
leads and dot. }\label{PIC.1}
\end{figure}

We assume that there are two electron reservoirs at chemical
potential, $\mu_L=\mu_R=\mu$, that are coupled to the dots via
tunnelling. We note, however, that our results are equally valid
in the case of only a single reservoir. Only a single empty
orbital energy level, $\varepsilon_i$, of the dot lies close to
$\mu$. Here, the subscript $i$ denotes the particular dot. The
Zeeman splitting between the two electron spin states is
$\Delta_i=\varepsilon_{i,\downarrow}-\varepsilon_{i,\uparrow}=g^{(i)}_x\mu_{B}B$
where $B$ is a static magnetic field along the x-axis that is
perpendicular to the growth direction (z). $\mu_B$ is the Bohr
magneton and $g^{(i)}_x$  is the electronic g-factor of the
$i^{th}$ dot along the direction of the magnetic field. The energy
levels satisfy
$\varepsilon_{i,\uparrow}=\varepsilon_i-\Delta_i/2<\mu<\varepsilon_{i,\downarrow}=\varepsilon_i+\Delta_i/2$
so that only spin up electrons can tunnel into the dot and only
spin down electrons can tunnel out. In the limit of very large
Coulomb blockade energy, only a single electron from the reservoir
can occupy the dot resulting in the bare Hamiltonian for the dots
and cavity field,
\begin{equation}
H_0=\hbar\omega_{cav}\hat{a}^{\dagger}\hat{a}+\sum_{i=1}^{N}\varepsilon_{i,\uparrow}\hat{c}^{\dagger}_{i,\uparrow}\hat{c}_{i,\uparrow}+\varepsilon_{i,\downarrow}\hat{c}^{\dagger}_{i,\downarrow}\hat{c}_{i,\downarrow}
\label{H0}
\end{equation}
where $\hat{c}_{_i,\sigma}(\hat{c}^{\dagger}_{i,\sigma})$ are
annihilation (creation) operators for electrons in dot $i$ with
spin $\sigma$ in the x-direction of the magnetic field.

Transitions between the different electronic Zeeman states of a
dot are induced via a two-photon Raman transition involving a
strong laser field that may be treated classically and a quantized
mode of the microcavity \cite{Djuric-Search-1,imamoglu}. The two
optical fields couple the electron spin states to a higher energy
charged exciton state (trion) \cite{atature,chen,greilich, dutt}.
The lowest energy trion states excited by $\sigma^{+}$ and
$\sigma^{-}$ circularly polarized light are
$|+t_i\rangle=\hat{c}^{\dagger}_{i,\uparrow}\hat{c}^{\dagger}_{i,\downarrow}\hat{h}^{\dagger}_{i,+3/2}|0\rangle$
and
$|-t_i\rangle=\hat{c}^{\dagger}_{i,\uparrow}\hat{c}^{\dagger}_{i,\downarrow}\hat{h}^{\dagger}_{i,-3/2}|0\rangle$
, respectively. They consist of an electron singlet state and a
heavy hole, where $\hat{h}^{\dagger}_{i,\pm 3/2}$ are heavy hole
creation operators with spin projections $\pm \hbar 3/2$ along the
z-axis and $|0\rangle$ is the empty dot state. The $\sigma^{+}$
polarized pump laser with frequency $\omega_l$ and Rabi frequency
$\Omega_l$ couples each of the electron spin states to the
$|+t_i\rangle$ trion state. Similarly, we assume that the cavity
field, with vacuum Rabi frequency $g_{cav,i}$ for each dot and
frequency $\omega_{cav}$, is also $\sigma^{+}$ polarized due to
either the cavity construction or because it is driven by a
$\sigma^{+}$ pump as discussed below \cite{note}. When the two
fields are far detuned from the creation energy for the
$|+t_i\rangle$ state, the intermediate trion state can be
adiabatically eliminated to give
\begin{equation}
H_I=i\hbar
\sum_{i=1}^{N}g_i(\hat{a}\hat{c}^{\dagger}_{i,\downarrow}\hat{c}_{i,\uparrow}e^{i\omega_lt}-h.c.).
\label{HI}
\end{equation}
where $g_i=g_{cav,i}\Omega_l/4\Delta_R$ and $\Delta_R$ is the
detuning from the trion state. In deriving Eq. \ref{HI}, we have
assumed that $|\Delta_i- (\omega_{cav}-\omega_l)|\ll
|\Delta_i+(\omega_{cav}-\omega_l)| $ so that the non-resonant
terms
$\hat{a}^{\dagger}\hat{c}^{\dagger}_{i,\downarrow}\hat{c}_{i,\uparrow}e^{-i\omega_lt}+h.c.$
can be neglected. Since electrons enter the dot in the spin
$\uparrow$ state, a photon must be absorbed from the cavity mode
in order to generate a spin current. It is therefore necessary to
pump the cavity field. We assume that the cavity is driven by a
classical source oscillating at frequency $\omega_p$,
\begin{equation}
H_{P}=i\hbar \epsilon (\hat{a}^{\dagger}e^{-i\omega_pt}-h.c.),
\end{equation}
which generates a coherent state in an empty cavity
\cite{armen,savage,walls-milburn}.

The explicit time dependence can be removed from $H_{I}$ and
$H_{P}$ by transforming to a rotating frame for the field
operators, $\hat{a}=\hat{A}e^{-i\omega_pt}$ and the dot operators,
$\hat{c}_{i,\uparrow}=\hat{C}_{i,\uparrow}\exp(-i(\omega_l-\omega_p)t/2)$
and
$\hat{c}_{i,\downarrow}=\hat{C}_{i,\downarrow}\exp(i(\omega_l-\omega_p)t/2)$.
The Hamiltonian in the rotating frame is $H'=H'_0+H'_P+H'_I$,
\begin{widetext}
\begin{eqnarray}
H'_0&=&\hbar(\omega_{cav}-\omega_p)\hat{A}^{\dagger}\hat{A}+\sum_{i=1}^{N}\left(\varepsilon_i(\hat{C}^{\dagger}_{i,\uparrow}\hat{C}_{i,\uparrow}+\hat{C}^{\dagger}_{i,\downarrow}\hat{C}_{i,\downarrow})+
(\Delta_i+\omega_l-\omega_p)(\hat{C}^{\dagger}_{i,\downarrow}\hat{C}_{i,\downarrow}-\hat{C}^{\dagger}_{i,\uparrow}\hat{C}_{i,\uparrow})/2
\right) \label{H0'}\\
H'_I+H'_P &=&
i\hbar\sum_{i=1}^{N}g_i(\hat{A}\hat{C}^{\dagger}_{i,\downarrow}\hat{C}_{i,\uparrow}-h.c.)+
i\hbar \epsilon(\hat{A}^{\dagger}-h.c.)
\end{eqnarray}
\end{widetext}
From Eq. \ref{H0'}, one can clearly see that the resonance
conditions are $\omega_{cav}=\omega_p$ and
$\Delta_i=\omega_p-\omega_l$. We assume that former condition is
always satisfied, while the latter condition cannot be satisfied
for all dots due to variations in the magnetic moments between
dots.

The dynamics of the system can be described in terms of the
density operator, $\rho$ for the cavity plus dots. The master
equation for $\rho$ is given by,
\begin{equation}\dot{\rho}=-i[H',\rho]-\Gamma_{cav}(\hat{A}^{\dagger}\hat{A}\rho
-2\hat{A}\rho\hat{A}^{\dagger}+\rho\hat{A}^{\dagger}\hat{A})/2+\dot{\rho}|_{lead}
\label{master}
\end{equation}
The first term describes coherent dynamics of the coupled
QD-cavity system, the second term stands for the cavity decay
\cite{walls-milburn}, and the third term describes QD-lead
coupling. Here we assume that the Coulomb blockade is so large
that a second electron cannot tunnel into the dot if there is
already one electron in the dot. The lead-dot coupling is most
easily expressed in terms of the matrix elements of the density
operator, $\rho^{(n,m)}_{\sigma_i,\sigma'_i}=\langle
n,\sigma_i|\rho|\sigma'_i,m\rangle$ where $|\sigma_i,n\rangle$
represents a state with $n$ photons in the cavity and
$\sigma_i=0,\uparrow,\downarrow$ corresponding to no electrons,
one spin up, or one spin down electron, respectively, on the
$i^{th}$ dot. The specific form of the master equations for the
lead coupling are \cite{dong,Djuric-Search-1}
\begin{eqnarray}
\dot{\rho}^{(n,m)}_{0_i,0_i}|_{lead}&=& \Gamma^{(-)}_{i}
\rho^{(n,m)}_{\downarrow_i,\downarrow_i}-\Gamma^{(+)}_i\rho^{(n,m)}_{0_i,0_i}
\label{lead1}
\\
\dot{\rho}^{(n,m)}_{\uparrow_i,\uparrow_i}|_{lead}&=&\Gamma^{(+)}_i\rho_{0_i,0_i}^{(n,m)}
\\
\dot{\rho}^{(n,m)}_{\downarrow_i,\downarrow_i}|_{lead}&=&-\Gamma^{(-)}_i\rho^{(n,m)}_{\downarrow_i,\downarrow_i}
\\
\dot{\rho}^{(n,m)}_{\uparrow_i,\downarrow_i}|_{lead}&=&-\Gamma^{(-)}_i\rho^{(n,m)}_{\uparrow_i,\downarrow_i}/2.
\label{lead2}
\end{eqnarray}
Here, $\Gamma_i^{(-)}=\Gamma^{(-)}_{i,L}+\Gamma^{(-)}_{i,R}$ is
the rate at which spin down electrons tunnel out of the dots into
the left and right leads and
$\Gamma_i^{(+)}=\Gamma^{(+)}_{i,L}+\Gamma^{(+)}_{i,R}$ is the rate
at which spin up electrons tunnel into the dots. (The subscripts
$L$ and $R$ denote the tunnelling rates for the left and right
leads, respectively.) We also assume that the coupling between the
left and right leads and the dots are the same and that the
tunnelling between the leads and the dot is spin independent,
$\Gamma^{(+)}_{i, L}=\Gamma^{(-)}_{i,
L}=\Gamma^{(+)}_{i,R}=\Gamma^{(-)}_{i,R}=\Gamma$. It is worth
pointing out that because Eqs. \ref{lead1}-\ref{lead2} take into
account large Coulomb blockade in the dots, they are slightly
different from other master equations for dots that assume
noninteracting electrons \cite{sun}.

\section{Single Quantum Dot}
First we consider the limit of a single quantum dot and exact
two-photon resonance, $\Delta=\omega_{cav}-\omega_l$. We
numerically solve for the steady state density matrix by first
expressing the density matrix in vector form,
$\rho^{(n,m)}_{\sigma_i,\sigma'_i} \rightarrow \vec{\rho}$, and
rewriting Eq. \ref{master} in matrix form,
\begin{equation}
d \vec{\rho}/dt=M\vec{\rho}
\end{equation}
The steady state solution, $\vec{\rho}^{(0)}$, is given by the
eigenvector of $M$ with zero eigenvalue \cite{djuric}.

The steady state behavior of the system can be characterized by
the Q-distribution for the intracavity optical
field\cite{Meystre,walls-milburn},
$Q(\alpha)=\sum_{\sigma=0,\uparrow,\downarrow}\langle\alpha,\sigma|\rho^{(0)}|\alpha,\sigma\rangle/\pi$
where $|\alpha\rangle$ is a coherent state,
$\hat{a}|\alpha\rangle=\alpha|\alpha\rangle$. The advantage of the
Q-distribution is that it is positive semi-definite and can be
used to make comparisons to classical phase space probability
distributions. Figures 2 and 3 show $Q(\alpha)$ as well as the
probability distribution for the photon number, $P(n)$. One can
clearly see two peaks corresponding to the two stationary average
cavity field amplitudes.

\begin{figure}[htb]
\includegraphics[height=2.5in,width=3.5in]{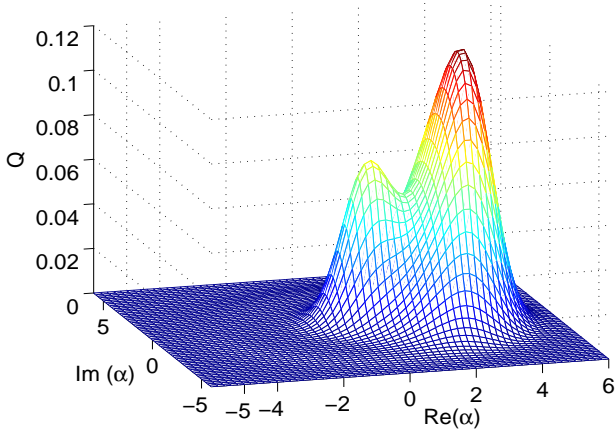}
\caption{Q-distribution vs. $Re[\alpha]$ and $Im[\alpha]$ for
$\Gamma_{cav}=0.2\Gamma$, $g=1.4\Gamma$ and
$\epsilon=0.5\Gamma$.}\label{FIG.1a}
\end{figure}

\begin{figure}[htb]
\includegraphics[height=2.0in,width=3.5in]{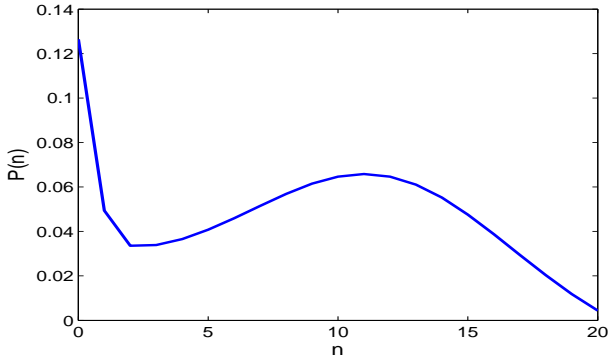}
\caption{Photon probability distribution, $P(n)$, vs. cavity field
photon number, $n$, for same parameters as Fig. 2,
$\Gamma_{cav}=0.2\Gamma$, $g=1.4\Gamma$ and
$\epsilon=0.5\Gamma$}\label{FIG.1b}
\end{figure}

The average spin current, $I_{s}=s(I_{\uparrow}-I_{\downarrow})$,
is the same in the left and right leads with the stationary
currents given by $\langle
I_{\uparrow}\rangle=\Gamma\rho^{(0)}_{0,0}$ and $\langle
I_{\downarrow}\rangle=-\Gamma\rho^{(0)}_{\downarrow,\downarrow}$.
Here $\rho^{(0)}_{\sigma,\sigma}$ are the populations of the dot
after tracing over the state of the cavity field. The average spin
current can be expressed in terms of expectation values of the
cavity field using Eq. \ref{master},
\begin{equation}
\langle I_S\rangle =2s(2\epsilon Re[\langle \hat{a} \rangle]-
\Gamma_{cav}\langle \hat{a}^{\dagger}\hat{a}\rangle)
\label{avspincurrent}
\end{equation}
Since the bimodality in $Q(\alpha)$ is in the amplitude,
$|\alpha|$, both $Re[\hat{a}]$ and $\hat{a}^{\dagger}\hat{a}$ have
two most probable values corresponding to the two peak locations,
$\alpha_{j}$ for $j=1,2$. Consequently, {\em individual}
measurements of the spin current would give results clustered
around two most probable values
\begin{equation}
I_{S,j}=2s(2\epsilon Re[\alpha_j]-
\Gamma_{cav}\alpha^*_j\alpha_j).
\end{equation}
By contrast, the ensemble averaged spin current of Eq.
\ref{avspincurrent} is $\langle
I_S\rangle=Tr[I_S\rho^{(0)}]\approx P_1 I_{S,1}+P_2 I_{S,2}$ where
$P_j$ are the total integrated probabilities for the two peaks in
$Q(\alpha)$. Because of the large quantum fluctuations associated
with a single dot, $\alpha^{(j)}$ are not truly stable points of
the system. Instead, this system should behave in a similar manner
to individual atoms in optical cavities, which exhibit stochastic
jumps between the two stationary cavity field values induced by
quantum noise \cite{armen,savage,Carmichael,Mabuchi2}.

\section{Semiclasical Bistability} Here we consider the equations
of motion for the field and dot expectation values,
\begin{eqnarray}
\alpha&=&\langle\hat{A}\rangle \\
s_i&=&\langle
\hat{C}^{\dagger}_{i,\uparrow}\hat{C}_{i,\downarrow}\rangle
\\
p_{i,\downarrow}&=&\langle\hat{C}_{i,\downarrow}^{\dagger}\hat{C}_{i,\downarrow}\rangle
\\
p_{i,\uparrow}&=&\langle\hat{C}^{\dagger}_{i,\uparrow}\hat{C}_{i,\uparrow}\rangle,
\end{eqnarray}
which can be derived from Eq. \ref{master} using
$\langle\dot{O}\rangle=Tr[O\dot{\rho}]$. In order to have a finite
system of equations, we must factorize the expectation values
involving the cavity field and dots,
$\langle\hat{A}^{\dagger}\hat{C}^{\dagger}_{i,\uparrow}\hat{C}_{i,\downarrow}\rangle\rightarrow\langle\hat{A}^{\dagger}\rangle\langle\hat{C}^{\dagger}_{i,\uparrow}\hat{C}_{i,\downarrow}\rangle
=\alpha^*s_i$. The resulting equations are then,
\begin{eqnarray}
\dot{p}_{i,\downarrow}&=&g_i(s_i\alpha^{*}+s_i^{*}\alpha)-\Gamma
p_{i,\downarrow} \label{semicl1}\\
\dot{p}_{i,\uparrow}&=&-g_i(s_i\alpha^{*}+s_i^{*}\alpha)+\Gamma
(1-p_{i,\downarrow}-p_{i,\uparrow}) \\
\dot{\alpha}&=& -\sum_{i=1}^{N} g_is_i-\Gamma_{cav}\alpha/2+\epsilon  \label{cavity-field}\\
\dot{s_i}&=& -i\delta_i s_i +
g_i\alpha(p_{i,\uparrow}-p_{i,\downarrow})-\Gamma s_i/2
\label{semicl}
\end{eqnarray}
where $\delta_i=\Delta_i+\omega_l-\omega_p$. This 'semiclassical'
factorization ansatz amounts to neglecting quantum mechanical
correlations between the cavity field and dots and is usually
assumed to be valid in the limit of large `classical'
systems\cite{armen}. Experiments with atoms in optical cavities in
the strong coupling regime showed good agreement with such a
semiclassical theory for $N>15$ atoms except for very close to the
end points of the bistable region\cite{rempe}. We therefore
restrict our treatment to $N\gg 1$ dots.

In order to derive analytic expressions, we will first neglect
inhomogeneous broadening, $\delta_i=0$, and variations of the
vacuum Rabi frequency due to random positions of the dots relative
to the cavity mode, $g_i=g_0$. We will account for these effects
later by numerical averaging over $\delta_i$ and $g_i$. We then
define new variables for the total population and polarization,
$P_{\sigma}=\sum_i p_{i,\sigma}$ and $S=\sum_i s_{i}$. By
introducing the polar representation, $S=|S|e^{i\theta}$,
$\alpha=|\alpha|e^{i\varphi}$ and $\epsilon=|\epsilon|e^{i\phi}$,
and using the positive definiteness of $P_{\sigma}$, one sees that
the phases are locked, $\varphi=\phi=\theta$.

The differential equations for $|\alpha|$, $|S|$, and $P_{\sigma}$
have steady state solutions given by
\begin{eqnarray}
12g_0^2\Gamma_{cav}|\alpha|_j^3&-&24g_0^2|\epsilon||\alpha|_j^2+
(4g_0^2\Gamma N+\Gamma^2\Gamma_{cav})|\alpha|_j \nonumber \\
&-&2\Gamma^2|\epsilon|=0
\label{alpha} \\
|S|_j&=&\frac{2|\epsilon|-\Gamma_{cav}|\alpha|_j}{2g_0} \\
P_{\downarrow j}&=&
\frac{-\Gamma_{cav}|\alpha|_j^2+2|\epsilon||\alpha|_j}{\Gamma} \\
P_{\uparrow j}&=&N-2P_{\downarrow j}.
\end{eqnarray}
Here, $|\alpha|_j$ for $j=1,2,3$, are the roots of Eq.~
\ref{alpha} and the steady spin current is given by
$I_{sj}=2s\Gamma P_{\downarrow j}$

In the limit of negligibly small cavity decay, $\Gamma_{cav}\ll
|\epsilon|$, $\Gamma$, $g_0$, Eq. ~(\ref{alpha}) becomes quadratic
with the two stationary solutions,
\begin{equation}
|\alpha|_{1,2}=\frac{\Gamma N}{12|\epsilon|}\left(
1\pm\sqrt{1-\frac{12|\epsilon|^2}{N^2g_0^2}}\right)
\label{absorptive}
\end{equation}
for $0<12|\epsilon|^2/N^2g_0^2<1$. By contrast, in the limit
$\Gamma\rightarrow 0$, the only nonzero solution for the cavity
field is $|\alpha|=2|\epsilon|/\Gamma_{cav}$. This limit is
different from the case of atomic OB where bistability in the
phase of $\alpha$ exists in the limit that the spontaneous
emission rate goes to zero \cite{Mabuchi2,Carmichael}. The
difference arises from the fact that $\Gamma$ represents both the
pumping rate and the decay rate for electrons in the dots. When
$\Gamma=0$, electrons cannot tunnel into the dots and interact
with the cavity, which effectively decouples the dots from the
cavity field.

\begin{figure}[htb]
\includegraphics[height=2.5in,width=3.5in]{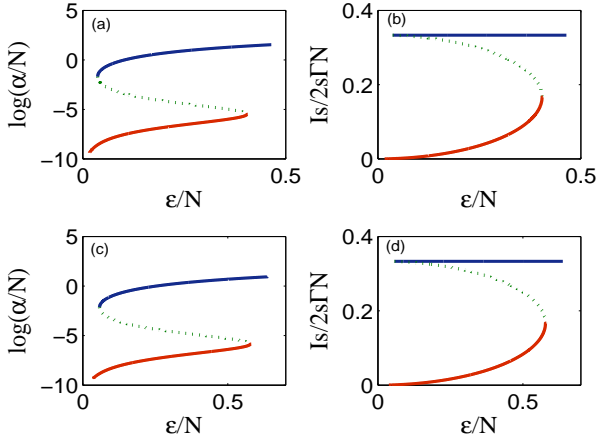}
\caption{(a,c) Cavity field amplitude $\log_{10}(|\alpha|/N)$ vs.
$|\epsilon|/N$.(b, d) Spin current, $I_S/2s\Gamma N$ vs.
$|\epsilon|/N$. For (a) and (b), $g_0=1.4\Gamma$ and
$\Gamma_{cav}=0.2\Gamma$. For (c) and (d), $g_0=2.0\Gamma$,
$\Gamma_{cav}=0.5\Gamma$. In all cases the number of dots is
$N=50$ and $|\epsilon|/N$ is plotted in units of $\Gamma$.
}\label{FIG.2}
\end{figure}

Solutions of Eq. \ref{alpha} are presented in Fig. 4 as a function
of the driving field, $\epsilon$. Three positive real solutions of
Eq. \ref{alpha} exist for $\epsilon$ in the interval
$|\epsilon|_1<|\epsilon|<|\epsilon|_2$, where
\begin{equation}
|\epsilon|_{1,2}=\frac{g_0N}{2\sqrt{6}}\sqrt{1+\frac{5}{2}\frac{N_0}{N}-\frac{1}{8}\left(\frac{N_0}{N}\right)^2\mp
\left(1-\frac{N_0}{N}\right)^{3/2}} \label{bistableregion}
\end{equation}
where $N_0=2\Gamma\Gamma_{cav}/g_0^2$ is the critical dot number.
By analogy to atomic cavity QED, it is the number of dots
necessary to significantly modify the resonant properties of the
cavity \cite{armen}. The requirement that $(1-N_0/N)^{3/2}$ be
real implies that $N_0<N$, which reduces to the requirement of
strong coupling to the cavity when $N=1$, as one would expect. Eq.
\ref{bistableregion} corresponds to the range of $|\epsilon|$ for
which bistability occurs. One can clearly see from Fig. 4 that the
bistability of $\alpha$ also leads to bistability and hysteresis
for the spin current. We have analyzed the stability of the three
steady state solutions as outlined in the appendix and determined
that the two positive slope solutions in Fig. 4 (represented by
the solid lines) are stable while the negative slope solutions
(dotted lines) are unstable. For the cavity field, the lower
stable branch corresponds to a nearly empty cavity, $\alpha\approx
0$ but with a {\em finite} increasing spin current, $I_S\approx 0$
up to $I_S \approx 2sN\Gamma\times0.2$. Since $g_0,\Gamma >
\Gamma_{cav}$, this solution corresponds to a photon entering the
cavity and being almost instantaneously absorbed by a dot, which
then leads to the creation of one unit of spin current. The upper
stable branch corresponds to a large cavity field, $\alpha \approx
0.1N$ up to $\alpha \approx 10N$, with a {\em constant} spin
current of $I_S=2s\Gamma N/3$. This value of the spin current
corresponds to a saturated transition in the dot with equal
probabilities for the two spin states along with the empty dot
state,
$\rho^{(0)}_{0_i,0_i}=\rho^{(0)}_{\uparrow_i,\uparrow_i}=\rho^{(0)}_{\downarrow_i,\downarrow_i}=1/3$.
In this case, a spin current per dot of $2s\Gamma/3$ was
previously found to be the maximum spin current that could be
generated by a single dot using a classical field \cite{dong}.

In order to account for variations in $\delta_i$ and $g_i$ as a
result of the random sizes and positions of the dots during the
growth process, we assume that the variations can be modelled
using Gaussian probability distributions,
\begin{eqnarray}
P_d(\delta_i)&=&\frac{1}{\sigma_d\sqrt{2\pi}}e^{-\delta_i^2/2\sigma_{d}^2}
\\
P_g(g_i)&=&\frac{1}{\sigma_g\sqrt{2\pi}}e^{-(g_i-g_0)^2/2\sigma_{g}^2}.
\end{eqnarray}
The contribution of each of the dots to the cavity field appears
as a summation over the dot polarizations in Eq.
\ref{cavity-field}, which must now be replaced with an integration
over the probability distributions, $\sum_i g_is_i\rightarrow
N\int\int d\delta_i dg_i P_d(\delta_i)P_g(g_i) g_is_i$. The
resulting equation for the steady state cavity field is then
\begin{eqnarray}
|\epsilon|-\Gamma_{cav}|\alpha|/2 &=& N\int d\delta_i dg_i
P_d(\delta_i)P_g(g_i) f(\delta_i,g_i) \label{inhomogoneous} \\
f(\delta_i,g_i)&=&
\frac{g_i^2|\alpha|\Gamma}{6g_i^2|\alpha|^2+\Gamma^2/2+2\delta_i^2}
. \label{inhomogeneous-2}
\end{eqnarray}
The cavity field amplitude can then be used to calculate the spin
current using $I_s=2s\Gamma \sum_i
p_{i,\downarrow}=2s(2|\alpha||\epsilon|-\Gamma_{cav}|\alpha|^2)$.
Note that in deriving Eq. \ref{inhomogoneous} we have made use of
the fact that the sine of the phase difference between the cavity
and driving fields, $\sin(\phi-\varphi)$, vanishes when averaged
over a probability distribution that is even about $\delta_i=0$.
One can see from Eq. \ref{inhomogoneous}-\ref{inhomogeneous-2},
that for small cavity fields $g_i|\alpha|\ll \Gamma$, the
homogenous broadening of the dot levels due to the leads dominates
over the inhomogeneous broadening of the dot levels when $\sigma_d
\ll \Gamma$. On the other hand for large cavity fields,
$g_i|\alpha| \gg \Gamma,\sigma_d$, power broadening dominates over
homogeneous and inhomogeneous broadening of the dot levels and the
cavity field is independent of $\sigma_d$.

Figures 5 and 6 show the cavity field amplitude and spin current
as a function of the driving field amplitude, $\epsilon$, for
different $\sigma_d$ and $\sigma_g$. For the Rabi frequencies,
$\sigma_g$ was chosen to be $20\%$ or $40\%$ of the mean while for
the Zeeman splitting we chose $\sigma_d>\Gamma$ for all curves.
One can see that for increasing $\sigma_d$ and $\sigma_g$ the
range of $\epsilon$ where bistability occurs is reduced. This can
be understood as the result of fewer dots that are both resonant
with the cavity mode and strongly coupled to it.

\begin{figure}[htb]
\includegraphics[height=2.5in,width=3.5in]{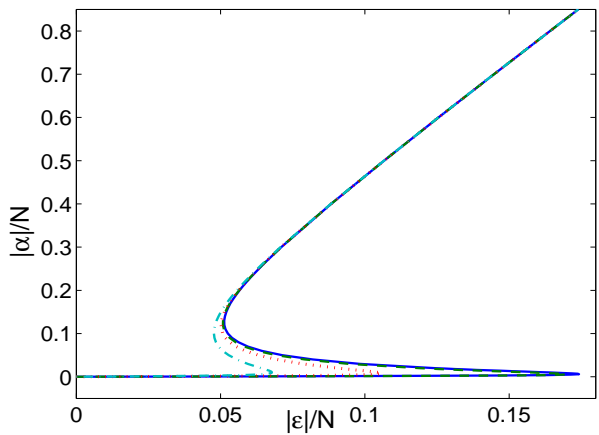}
\caption{Cavity field amplitude, $|\alpha|/N$, vs. $|\epsilon|/N$
in units of $\Gamma$ for $N=50$ dots, $g_0=1.6\Gamma$, and
$\Gamma_{cav}=0.4\Gamma$. The different curves correspond to:
$\sigma_d=2\Gamma$ and $\sigma_g=0.2g_0$ (solid line);
$\sigma_d=2\Gamma$ and $\sigma_g=0.4g_0$ (dashed line);
$\sigma_d=4\Gamma$ and $\sigma_g=0.2g_0$ (dotted line);
$\sigma_d=7\Gamma$ and $\sigma_g=0.2g_0$ (dashed dot line).}
\end{figure}

\begin{figure}[htb]
\includegraphics[height=2.5in,width=3.5in]{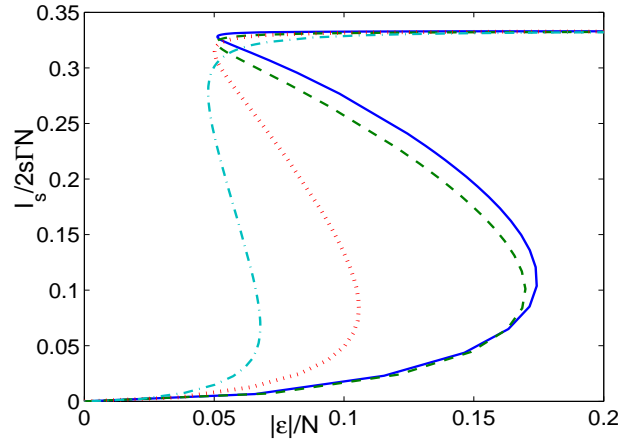}
\caption{Spin current, $I_s/2s\Gamma N$, vs. $|\epsilon|/N$ in
units of $\Gamma$ for the same parameters as Fig. 5: $N=50$ dots
and $g_0=1.6\Gamma$, $\Gamma_{cav}=0.4\Gamma$. The different
curves correspond to: $\sigma_d=2\Gamma$ and $\sigma_g=0.2g_0$
(solid line); $\sigma_d=2\Gamma$ and $\sigma_g=0.4g_0$ (dashed
line); $\sigma_d=4\Gamma$ and $\sigma_g=0.2g_0$ (dotted line);
$\sigma_d=7\Gamma$ and $\sigma_g=0.2g_0$ (dashed dot line).}
\end{figure}

\section{Conclusion}
In conclusion, we have studied the generation of spin currents by
electrical transport through quantum dots coupled to a single mode
of an optical microcavity. We have shown that when the cavity
field is coherently driven, absorptive optical bistability occurs
in the cavity field. Since the steady state spin current is
proportional to both the cavity field amplitude and number of
photons in the cavity, OB necessarily leads to bistability in the
spin current. Moreover, we have shown that this bistability
persists in the presence of inhomogeneous broadening of the dots'
Zeeman splitting and vacuum Rabi frequencies.

The cavity field and spin current could be made to switch between
the two stable states by varying the driving field amplitude,
$\epsilon$, beyond the endpoints of the bistable
region\cite{Meystre}. One might then envision using this system as
an optically controlled spin current switch, which could be used,
for example, to transfer optically encoded digital information
into a spin current. In a future publication we plan to explore
quantum noise induced switching between stationary spin currents
states for the single quantum dot case using a stochastic master
equation \cite{Mabuchi2}.

\section{Appendix: Linear Stability Analysis}
In order to analyze the stability of the steady state solutions,
we consider small fluctuations $\delta P_{\downarrow}$, $\delta
P_{\uparrow}$, $\delta\alpha$ and $\delta s$ about steady the
state solutions of Eqs. \ref{semicl1}-\ref{semicl},
\begin{eqnarray}
P_{\downarrow}(t)&=&P_{\downarrow}^{(0)}+\delta
P_{\downarrow}(t) \label{fluctuation1}\\
P_{\uparrow}(t)&=&P_{\uparrow}^{(0)}+\delta P_{\uparrow}(t) \\
\alpha(t) &=&\alpha^{(0)}+\delta\alpha(t) \\
s(t)&=&s^{(0)}+\delta s(t). \label{fluctuation2}
\end{eqnarray}
where $P_{\downarrow}^{(0)}$, $P_{\uparrow}^{(0)}$, $\alpha^{(0)}$
and $s^{(0)}$ correspond to one of the three steady solutions
solutions given by Eq. \ref{alpha}. By inserting Eqs.
\ref{fluctuation1}-\ref{fluctuation2} into Eqs.
\ref{semicl1}-\ref{semicl} and discarding terms that are quadratic
in the fluctuations, we obtain the following linear equations for
the fluctuations,
\begin{widetext}
 \begin{equation} \frac{d}{dt}\left(%
\begin{array}{c}
  \delta P_{\downarrow} \\
  \delta P_{\uparrow} \\
  \delta\alpha \\
  \delta\alpha^* \\
  \delta s \\
  \delta s^* \\
\end{array}%
\right)=\left(%
\begin{array}{cccccc}
  -\Gamma & -\Gamma & gs^{*(0)} & gs^{(0)} & g\alpha^{*(0)} & g\alpha^{(0)} \\
  0 & -\Gamma & -gs^{*(0)} & -gs^{(0)} & -g\alpha^{*(0)} & -g\alpha^{(0)} \\
  0 & 0 & -\frac{\Gamma_{cav}}{2} & 0 & g & 0 \\
  0 & 0 & 0 & -\frac{\Gamma_{cav}}{2} & 0 & g \\
  -g\alpha^{(0)} & g\alpha^{(0)} & g(P_{\uparrow}^{(0)}-P_{\downarrow}^{(0)}) & 0 & -\frac{\Gamma}{2} & 0 \\
  -g\alpha^{*(0)} & g\alpha^{*(0)} & 0 & g(P_{\uparrow}^{(0)}-P_{\downarrow}^{(0)}) & 0 & -\frac{\Gamma}{2} \\
\end{array}%
\right)\left(%
\begin{array}{c}
  \delta P_{\downarrow} \\
  \delta P_{\uparrow} \\
  \delta\alpha \\
  \delta\alpha^* \\
  \delta s \\
  \delta s^* \\
\end{array}%
\right)={\bf J}\left(%
\begin{array}{c}
  \delta P_{\downarrow} \\
  \delta P_{\uparrow} \\
  \delta\alpha \\
  \delta\alpha^* \\
  \delta s \\
  \delta s^* \\
\end{array}%
\right)
\end{equation}
\end{widetext}
The associated characteristic equation for the eigenvalues of
${\bf J}$ is a sixth order polynomial
\begin{equation}
\sum_{i=1}^{6}a_l\lambda^l=0\label{c.e}
\end{equation}
A particular steady state solution,
$\{P_{\downarrow}^{(0)},P_{\uparrow}^{(0)},\alpha^{(0)},s^{(0)}
\}$, is stable if $Re[\lambda]<0$ for {\em all} six of the
eigenvalues since any small noise induced fluctuation will then
decay away exponentially. Steady state solutions with at least one
eigenvalue satisfying  $Re[\lambda]>0$ are unstable since small
fluctuations will grow exponentially with time. Numerical
solutions for the eigenvalues for each of the three steady states
in Figs. 4-6 indicate that two of the roots are stable and one is
unstable. The stable roots correspond to the upper and lower
branches in Figs. 4-6 that have positive slope while the middle
branch with negative slope is unstable.


\begin{references}
\bibitem{note} A linearly polarized cavity will also suffice.
However it will also couple the Zeeman states to the $|-t\rangle$
trion state leading to additional AC stark shifts.
\end{references}
\end{document}